# A Semantic Enhanced Model for effective Spatial Information Retrieval

Adeyinka K. Akanbi[1], Olusanya Y. Agunbiade[2], Sadiq Kuti[3], Olumuyiwa J. Dehinbo[4]

[1]Institute of Science & Technology, Jawaharlal Nehru Technological University,
Hyderabad 500085, A.P, India.
akanbiadeyinka@hotmail.com

[2]Department of Computer Science and Engineering, Tshwane University of Technology,
Pretoria Campus, South Africa.
agunbiadeOY@tut.ac.za

[3]Department of Electronics & Communication Engineering, Jawaharlal Nehru Technological University,
Hyderabad 500085, A.P, India.
kutisadiq@gmail.com

[4]Department of Computer Science & Engineering, Tshwane University of Technology,
Pretoria Campus, South Africa.
dehinboOJ@tut.ac.za

***Abstract*— A lot of information on the web is geographically referenced. Discovering and retrieving this geographic information to satisfy various users needs across both open and distributed Spatial Data Infrastructures (SDI) poses eminent research challenges. However, this is mostly caused by semantic heterogeneity in user's query and lack of semantic referencing of the Geographic Information (GI) metadata. To addressing these challenges, this paper discusses ontology-based semantic enhanced model, which explicitly represents GI metadata, and provides linked RDF instances of each entity. The system focuses on semantic search, ontology, and efficient spatial information retrieval. In particular, an integrated model that uses specific domain information extraction to improve the searching and retrieval of ranked spatial search results.**

***Index Terms*— Spatial Indexing, Spatial Reasoning, Spatial Ontology, and Semantics**

## I. INTRODUCTION

The past half century has witnessed rapid advancement in various computing and Information Technology [1], [19]. These developments in Computing & IT technologies have revolutionized information processing & the sharing of large volumes of earth Geographic Information (GI) such as survey data, satellite imagery and maps which runs into billions of gigabytes of geospatial data. The large volumes of spatial data provide valuable resources to both the ordinary users & researchers for their various uses, and are accessible for retrieval through Digital Libraries - Geolibrary, GeoPortals, Spatial Data Infrastructures (SDI) and the Web.

In Information Retrieval, the major aim of providing access to information data is probabilistic. It is concerned with precise issues as to whether a document is relevant for a user and request. Whereas, data retrieval is deterministic with regard to retrieval operations. If a document fulfills the conditions specified in the user's query, then it is termed "relevant & appropriate". However, Spatial Information Retrieval, is concerned with both deterministic retrieval (such as finding all data sets that contain information regarding a particular entity) and probabilistic retrieval (such as finding spatial relationship among entities in a domain.) [30].

Currently, one of the most important challenges against effective Spatial Information Retrieval (SIR) is the lack of detailed sematic referencing showing the spatial relationship among entities of the geospatial data for easy accessibility by the user. According to [2], Spatial Information Retrieval is mainly concerned with the provision of access to geo-referenced information sources, with indexing and retrieval of spatially oriented information. Users of digital libraries need to be able to search for specific known items in the database and to retrieve relevant unknown items based on various criteria. This searching and retrieval has to be done efficiently and effectively, even when dealing with large scale of data [5], [24]. This implies that digital library objects must be properly indexed so that users can retrieve them by content.

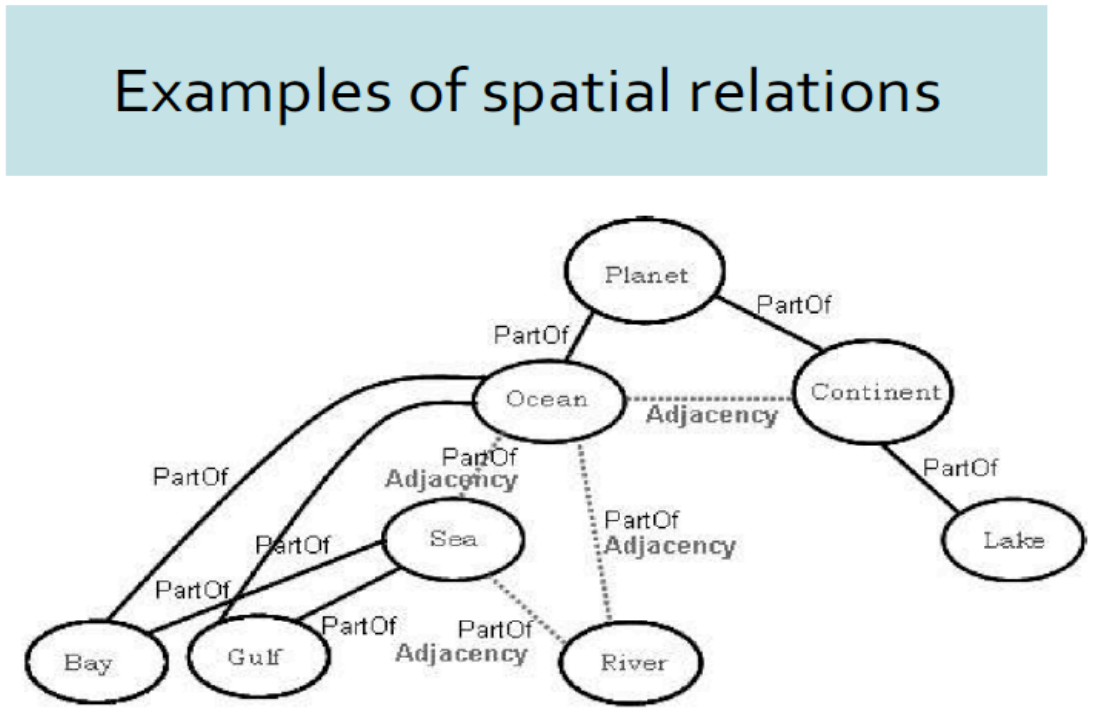

*Figure 1: Spatial relationship*

Indexing is required for both efficient access to large databases and to organize and limit the set of elements of a database that are accessible. Most Information Retrieval (IR) systems derive their index elements from the contents of the items to be indexed. The derivation may be simple extraction keywords from a text, inferential extraction such as mapping from text word to thesaurus terms, or it may be intellectual analysis and assignment of index items such as assigning subject headings to a document [30].

However, we envisage that for an effective SIR, there should be a need to improve the indexing of Geographic Information (GI), by representing the GI based on semantic ontology framework, which will provides a platform to showcase the common attributes such a location coordinates, area, data types and also provenance for identification based on sematic information containing the spatial relationships between the spatial entities in the domain. The lack of sematic model for referencing spatial data makes effective SIR impossible to accomplish.

This paper will examine the notion of Spatial Information Retrieval in the context of digital libraries; in particular, it will focus on the application of particular class of ontology for the semantic referencing of spatial information, and also effective indexing and retrieval methods appropriate for spatial information retrieval.

It should be also noted that, OpenGIS Consortium (OGC) has carried out far reaching research work on the retrieval of spatial information [35], and provides a specification that enables the syntactic interoperability and cataloguing of geographic information. This simple cataloguing system support discovering, organization, and access of geographic information [6], to some certain extent, however, they do not yet provide methods to solve problems of semantic heterogeneity, which is part of the problems mitigating against effective SIR. Problems of the semantic heterogeneity are mostly caused by synonyms and homonyms in metadata and user's query information [7]. In different part of the world, a single vocabulary term has different meaning; therefore solving this semantic heterogeneity is an important factor for achieving an efficient Spatial Information Retrieval.

A typical example, a user query for "Lodging Hotels in Hyderabad", if there is no explicit semantic referencing, indicating the difference between a lodging hotel and catering hotel the user will be presented with vague results of hotels that might not be providing lodging facilities for guests. This motivation example clearly shows the problems of heterogeneity caused by synonyms and homonyms due to lack of semantic referencing differentiating the different entities in the domain.

Considering these factors, it is necessary to use or develop a conceptual semantic ontological model for spatial data referencing. The most widely accepted common conceptualization of the geographic world is based on ideas of objects and fields [3], [4]. To this end, the possible approach to overcome the problems of semantic heterogeneity and enhance semantic referencing of GI metadata is the explication of knowledge, by means of ontology, which can be used for the identification and association of semantically corresponding concepts, because ontology can explicitly and formally represent concepts and relationships between concepts and can support semantic reasoning according to different entities in the domain [9].

Many definitions have been proposed for Ontology, such as: ontology is a formal, explicit specification of a shared conceptualization, by [9]. Ontologies are theories that use a specific vocabulary to describe entities, classes, properties, and functions related to a certain view of the world. They can be a simple taxonomy, a lexicon or a thesaurus, or even a fully axiomatized theory [11], which allows classification of different entities in the domain, by classes for easy semantic referencing and identification of the geographic information. Moreover, this improves the accuracy of searching and enables the development of powerful applications that execute complicated queries, whose answers do not reside on a single web page [10].

The Semantic Web relies heavily on the formal ontologies that structure underlying data for the purpose of comprehensive and transportable machine understanding [34]. Therefore, the success of the Semantic Web depends strongly on the proliferation of ontologies, which requires fast and easy engineering of ontologies and avoidance of a knowledge acquisition bottleneck. With semantic web technology, web information is given a well-defined meaning that can be understood by machines. The adoption of geographic ontology will surely enhance SIR.

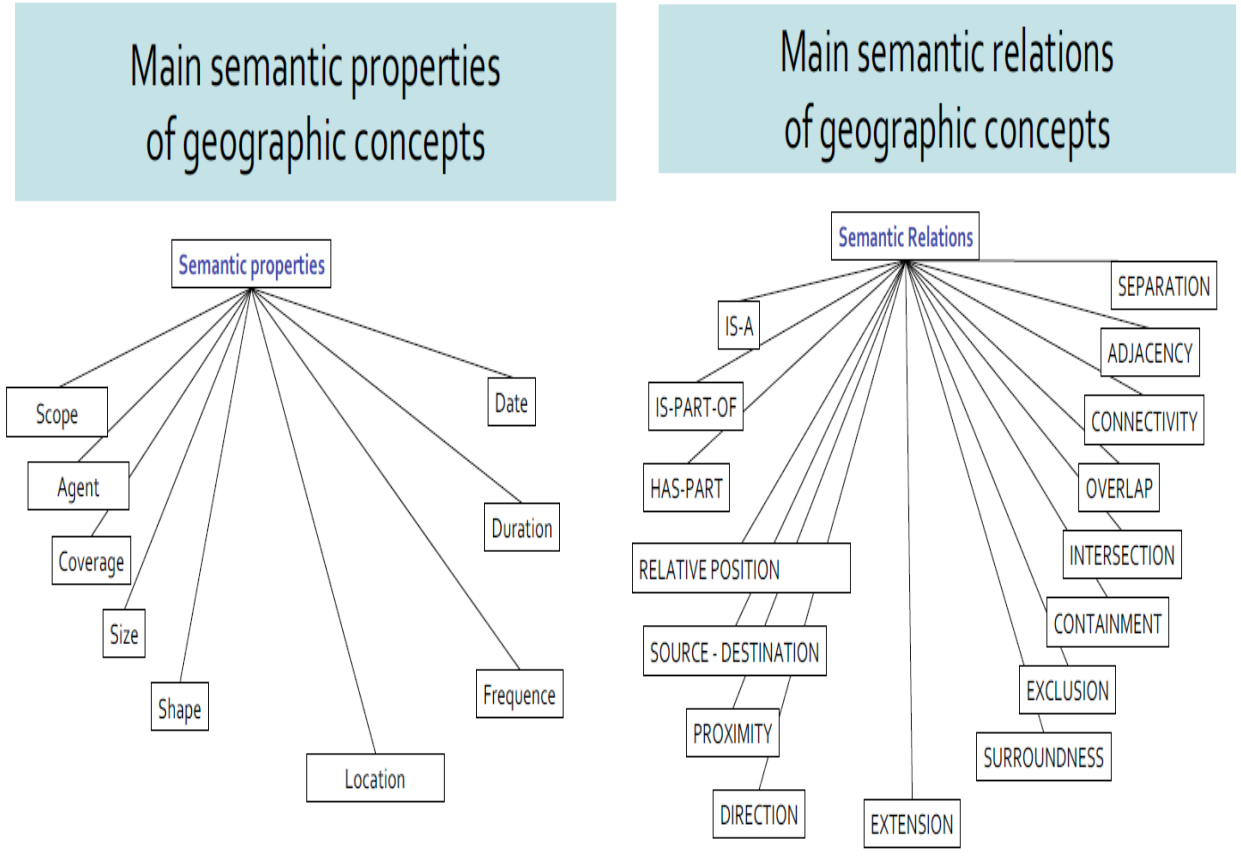

*Figure 2: Semantic properties and relations of geographical concepts.*

Therefore, we decided to up the ante, by adopting a universally Geographic Ontology (GO): GeoNames (GN), for semantic referencing of GI metadata, and not just a simple ontological framework [8] in solving the semantic referencing.

The goal of this paper is to ensure effective SIR, through the use of geographic ontology for the semantic referencing of spatial information and reduction of semantic heterogeneity for an effective SIR across all digital libraries. The ontology adopted is GeoNames Ontology: an ontology-based semantic description model to explicitly represent geographic information. Subsequent sections cover previous related work in SIR, define and describe the ontology-based approach for discovering and retrieval, indexing, searching, and spatial querying. Finally, we make conclusions and discuss the future work need to do.

## I. RELATED SYSTEMS

Spatial information retrieval (SIR) is mostly adopted in different kind of online and stand-alone applications. It is related to spatial queries that refer to location of objects, features of objects, or geographical information of a particular geographical entity. Semantic enhanced SIR is a fast developing area of research with various research work carried out e.g. [7],[12],[13].
Klien et al (2004) [7] presented a design for ontology-based discovery and retrieval of geographic information for solving existing problems of semantic heterogeneity. It adopts a simple catalogue based system for classification of entities.

SPIRIT (Spatially-Aware Information Retrieval on the Internet) [12] project developed tools and techniques to support spatial search on the Internet based on ontology [14], to assist spatial search [15].

Hartwig H. Hochmair (2005) [13] proposed a conceptual framework to overcome problems of semantic heterogeneity in keyword-based retrieval of geographic information. In the architecture, the server-sided knowledge base including domain ontology and rules for query expansion is used to expand the keyword-based searches [8].

A major shortcoming of these approaches is that they cannot recognize alternative names for the same place, whether they are literarily names or historical variants. In this paper, focus will lies on uniform semantic descriptions of geographic information based on GO: GeoNames, for discovery and retrieval based on semantic descriptions.

## II. SEMANTIC MODEL BASED ON ONTOLOGIES

Searching and retrieval of spatial information is performed by the execution of spatial queries based on user-input keywords. However, the execution of this keyword queries are not efficient enough for Spatial Information Retrieval due to the lack of semantics and inference mechanism [13]. The ability of adopting semantic referencing provides possibility to enhance Spatial Information Retrieval.

In Feb 2004, The World Wide Web Consortium released the Resource Description Framework (RDF) and the OWL Web Ontology Language (OWL) as W3C Recommendations. RDF is used to represent information and to exchange knowledge in the Web. OWL is used to publish and share sets of terms called ontologies, supporting advanced Web search, software agents and knowledge management [33].

The major problem with the Information retrieval of spatial data is also due to the inconsistency of metadata information. This apparent lack of detailed metadata information hampers the effective semantic referencing of spatial data, and can only be corrected through the adoption of a universally adopted Geographic Ontology. Ontology has also been regarded as an alternative to enhance the task of information retrieval [27]. In this section, we describe a semantic enhanced model that explicitly represent semantic of spatial information in a machine-readable format. Thus, all spatial data are semantically referenced and effective spatial information retrieval is achieved.

### ONTOLOGIES

Ontologies are the vital to semantic description & referencing of geographic information. They contain domain knowledge, specific data regarding a certain subject field, in a very structured way [16]. Thus, Geographic Ontology are ontologies with spatial relationships between geographic features. If an ontology is known and used by everybody for annotating information and searching for information, then all the above problems of search and retrieval are eliminated [17]. In this paper we adopted a modified version of GeoNames a widely used geographic ontology, due to efficient semantic referencing of geographic information metadata e.g. names, latitude, longitude and other information, and ability to provide linked RDF instances to various GI. In IR, ontology can be used on various levels [29], first, it makes it possible to refine a system based on a traditional process of indexing by increasing the chances to formulate a request starting from the terms or descriptors representing as well as possible information requirements. This process presents several interests: In current standards-based catalogues e.g. [22], users can formulate queries using keywords and/or spatial filters [20]. The metadata fields are predefined which are mostly based on the metadata schema used (e.g. ISO 19115) and on the query functionality of the service.

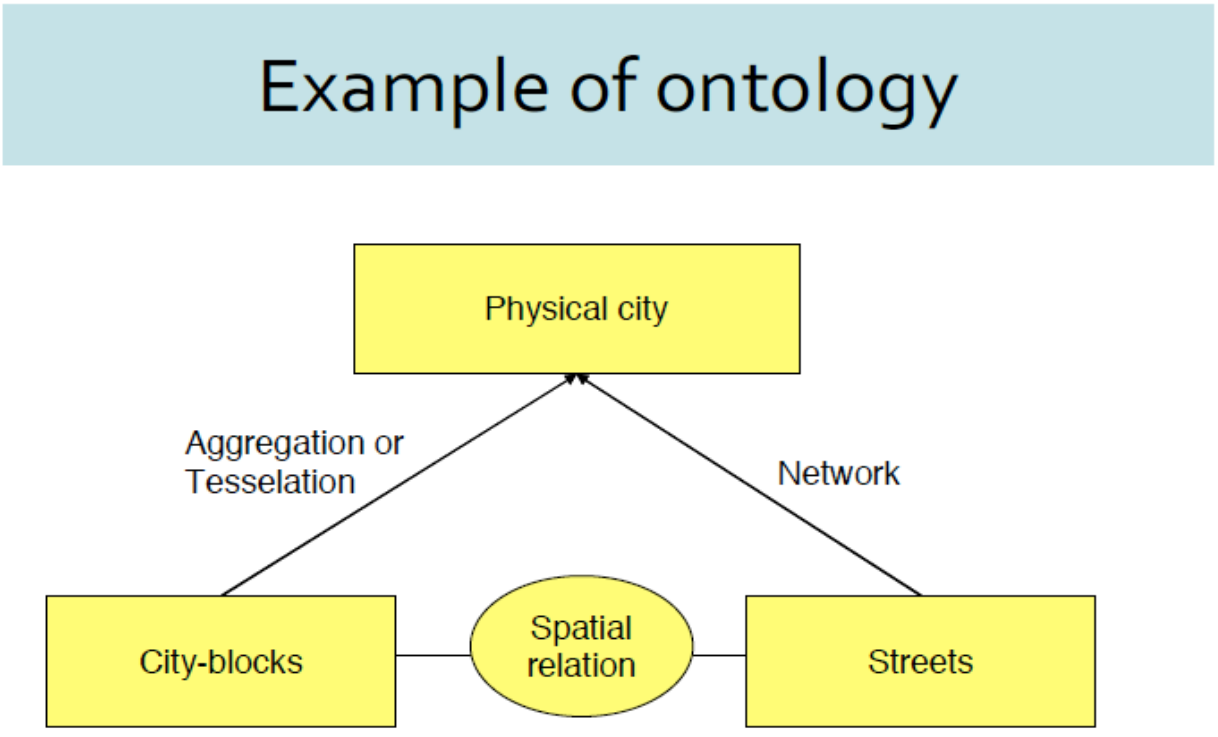

*Figure 3: Ontology*

## III. SPATIAL INFORMATION RETRIEVAL CONCEPTUAL FRAMEWORK

Solving the problems of semantic heterogeneity between user's search query and metadata description of geographic information is tantamount to an effective Spatial Information retrieval. The figure below, describe a SIR model for executing SIR queries, with various components such as s data source, spatial indexer, query engine, and ranker.

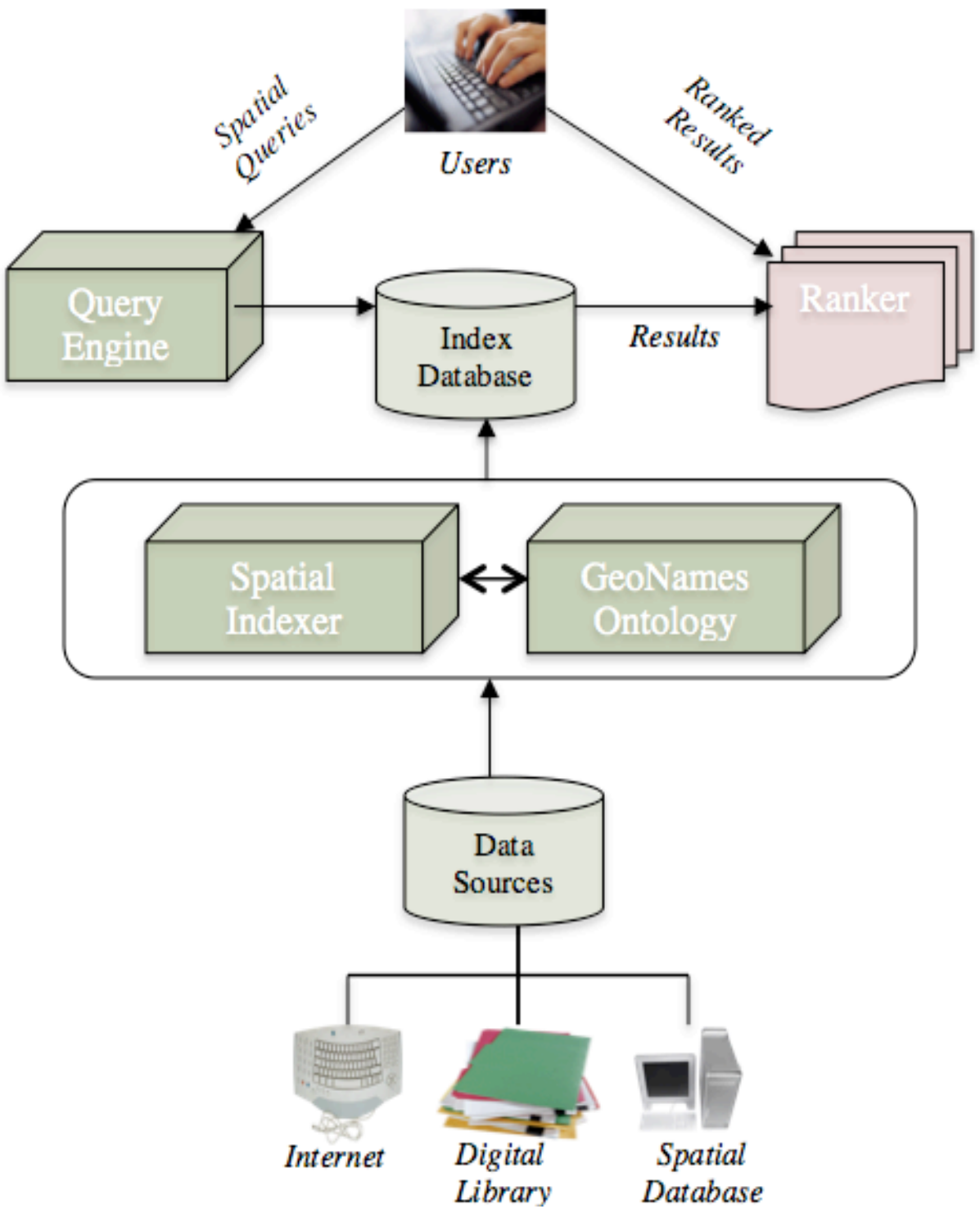

*Figure 4: Spatial Information Retrieval model.*

The Spatial indexer provides spatial indexing by extracting geographic locations & coordinates in a text or mapping data to a term based on the geospatial ontology; GeoNames [25, 26]. Query engine handles user requests [26]. To provide a better Quality of Service (QoS), a ranker is normally used to sort the results based on match level.

## A. SPATIAL QUERIES

Spatial queries simply imply querying a spatially indexed database based on relationships between different entities in that database within a particular domain. Spatial querying is the more general term. It can be defined as queries about the spatial relationships of entities geometrically defined and located in space [32]. In the following discussion, the emphasis will be on spatial queries; we will examine and concentrate on the basic types of spatial queries with regards to SIR.

### TYPES OF SPATIAL QUERIES

The various types of spatial queries submitted by the user to an information system such as SDIs may be arbitrarily complex in the types of information requested. Therefore, an effective SIR system should be able to execute spatial queries such as (1) Point-in-polygon query, might ask which satellite images are available that show a particular spot or which documents describe the place indicated by the point (denoted by [x, y]); (2) Region query, (3) Buffer Zone query refers to finding spatial objects that are within certain distance of an area; (4) Path Query, which requires the presence of a network structure to do shortest path or shortest time planning and (5) Multimedia query [2].

## B. INDEXING

Semantic index structure is the backbone in information retrieval performance. Retrieval of the spatial information is actually based on the user input spatial query. The spatial information retrieval system combines keyword-based querying of resources repository with the ability to query against RDF annotations of those resources based on the GeoNames Ontology.

RDF & RDF Schema are used to specify and populate ontology; the resulting RDF annotations are then indexed along with the full text of the annotated resources. The resultant index allows both keyword querying against the document full text and the literal values occurring in the RDF. This approach is considered as a key enabler for effective SIR. Furthermore, this approach holds that the ability to search and query the ontology, more fully supports typical information seeking and task annotation retrieval [31]. Traditional Information Retrieval techniques base their ranking algorithms on keyword weighting; our approach relies on the semantic concepts of ontology and indexed RDF annotations, for the ranking of query answers.

## C. SEARCHING & RETRIEVAL

Generally, the OWL format is used to store the extracted data in RDF, and SPAQRL or RDQL is used in querying with RDF language. However, accessing and extracting semantic spatial information from documents are beneficial to both people and machines. People can retrieve the documents semantically and machines can easily process the structured representation. OWL provides an expressive shared vocabulary to represent knowledge in the Semantic Web. The vocabulary allows expressing axioms about classes, properties, and individuals in the domain [28].

The application of ontology in the searching & retrieval of spatial information reduces the problem of semantic heterogeneity, which are classified by [21] as:

1. Naming heterogeneity (synonyms): This is due to lack of metadata description due to slightly different terminology.
2. Cognitive heterogeneity (homonyms): The user query tends to bring out result with little correlation to the intended request, thus indicating the occurrence of cognitive heterogeneity.

In traditional indexing techniques, keyword-based information are retrieved if they contain keywords specified by the user's query. However, much spatial information is missed due to their lack of accurate metadata or even though they do not contain the user specified keywords. This limitation can be addressed through the use ontology with semantic enhancement mechanism.

## IV. ARCHITECTURE FOR SEMANTIC ENHANCED SPATIAL INFORMATION RETRIEVAL

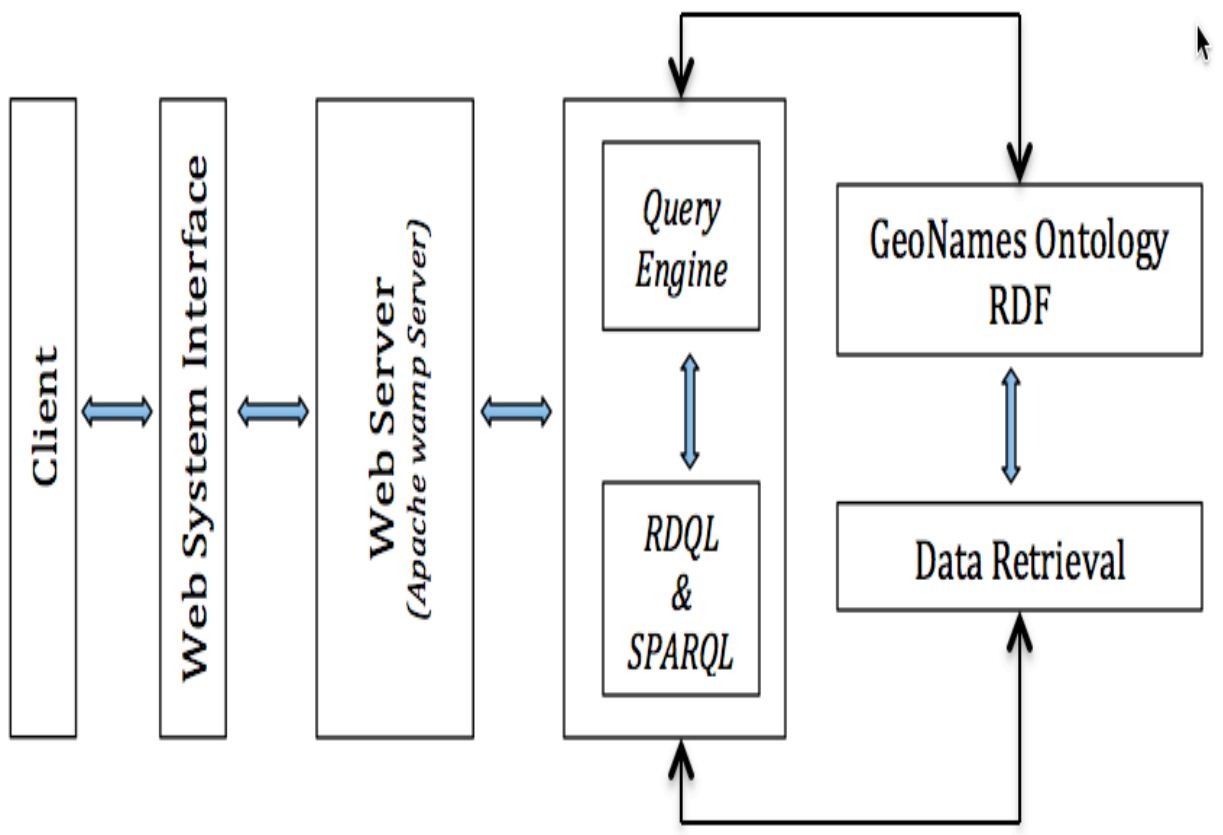

*Figure 5: System Architecture*

The Figure above presents the architecture of the system comprising several components, such as the client, user interface, web server, query engine, ontology repository, and data retrieval mechanism. The system was implemented by using various open source web technologies, such as Linux based wamp server etc. To query the index, the user enters a query into a Web interface and submits it to the Web server; the web server passes the spatial query along to the query engine (RDF Query Language generator), which retrieves the RDF instances based on the adopted ontology. The server processes the values of properties input by the client.

IMPLEMENTATION

The implementation of the above system architecture allows the spatial information retrieval via the ontology approach. The spatial information retrieval was implemented based on the adopted ontology providing accurate linked data to the RDF instances. The system helps the user to retrieve the spatial information and also helps in returning spatial information regarding the spatial query.

## V. PROPOSED SEMANTIC MODEL FOR SPATIAL INFORMATION RETRIEVAL

The proposed system aims to provide linked RDF data for the spatial query based on the indexed data adopted from the ontology framework.

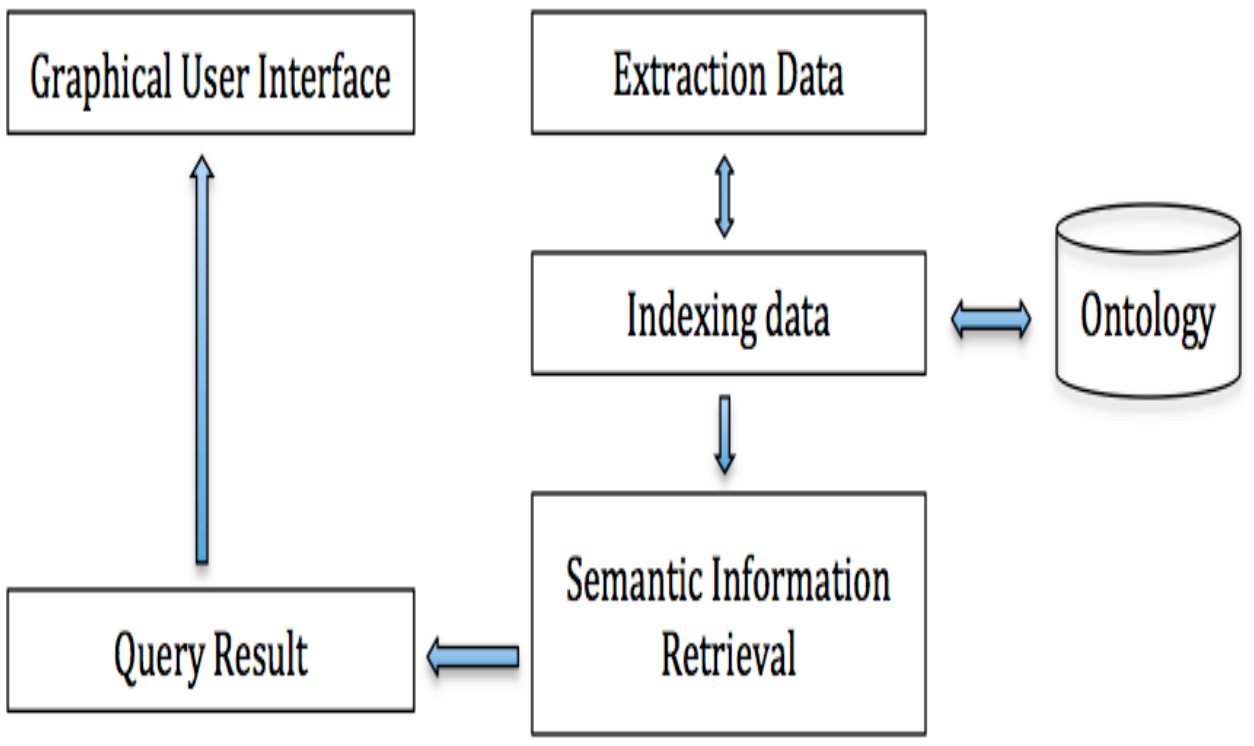

*Figure 6: Proposed semantic model based on an ontology framework*

### A. GEONAMES

Our model is based on a modified GeoNames Ontology, a GO where all countries and over eight million *placenames* are available in a geographical database, and offers RDF web services which is also part of a Linked Data project. As one of the most widely used geographical ontology, GeoNames Ontology makes it possible to add spatial semantic information to the Word Wide Web. The structure behind the data is the GeoNames ontology [18], which closely resembles the flat-file structure used in catalogue-based systems. Over 8.3 million GeoNames toponyms now have a unique URL with a corresponding RDF web service. Other services describe the relation between toponyms. A typical individual in the database is an instance of type Feature and has *aFeature* Class associated with it. *TheseFeature* Classes can be administrative divisions, populated places, structures, mountains, water bodies, etc. Though the *feature class* is subcategorized into 645 different *feature codes*, the *feature code* is associated with a *Feature instance* and not as a specialization of the property *featureClass*. A Feature also has several other properties, such latitude, longitude, and an *owl:sameAs* property linking it to an instance in the GeoNames database. This tends to increase the semantic referencing of the metadata. All coordinates use the World Geodetic System 1984 (WGS84) [23]

GeoNames is available as database dump, and also as Linked Open Data in the Resource Description Framework (RDF) format, meaning that each of the toponyms is identified by a public URI. E.g., information about the city Hyderabad will be found at http://sws.geonames.org/1269843/. Also categories or *feature codes* are defined in a formal OWL ontology and can be refried to by using a URI. Theses URIs can be used to add geographical information to other resources in the semantic web.

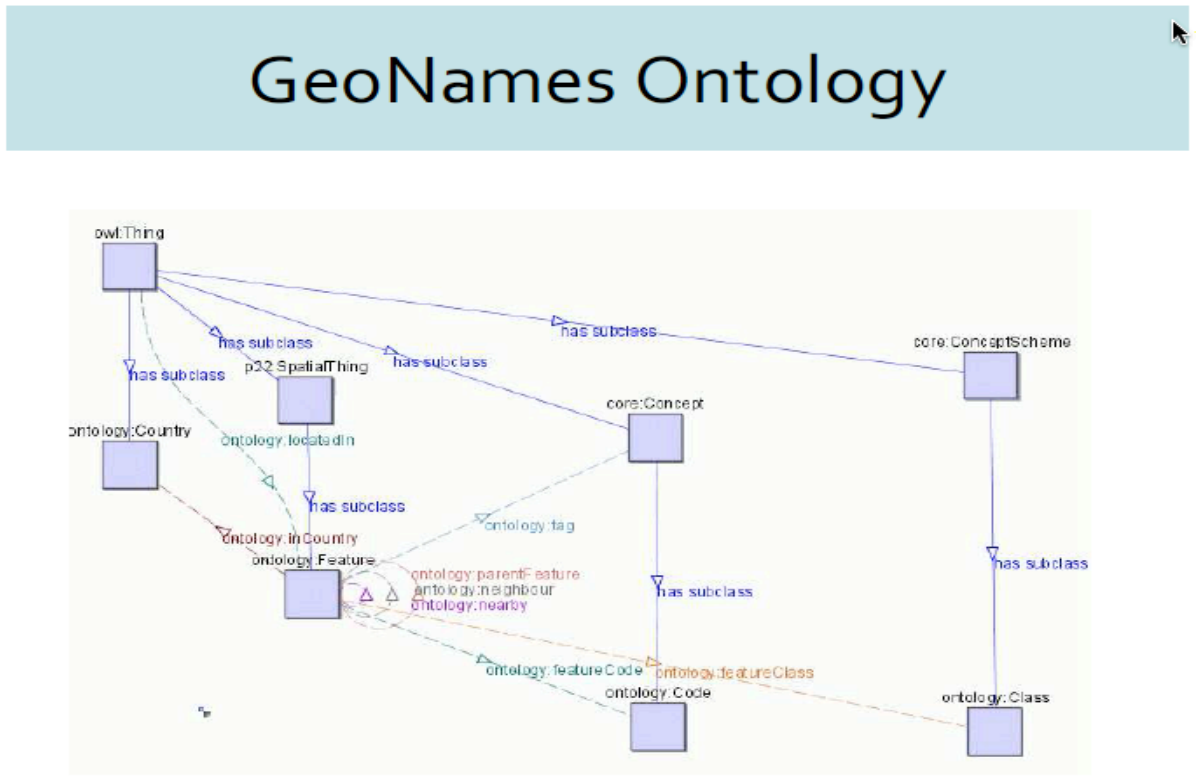

*Figure 7: GeoNames Ontology model*

The town Hyderabad in India we have these two URIs:
(*) http://sws.geonames.org/1269843/
(#) http://sws.geonames.org/1269843/about.rdf

The first URI (*) stands for the city in India. You use this URI if you want to refer to the city. The second URI (#) is the document with the information GeoNames has about Hyderabad. The GeoNames web server is configured to redirect requests for (*) to (#). The redirection tells Semantic Web Agents that Hyderabad is not residing on the GeoNames server but that GeoNames has information about it instead. Hence, an example of RDF description of a GeoNames *Feature* document, as obtained through the RDF Web service at URI http://sws.geonames.org/1269843/about.rdf

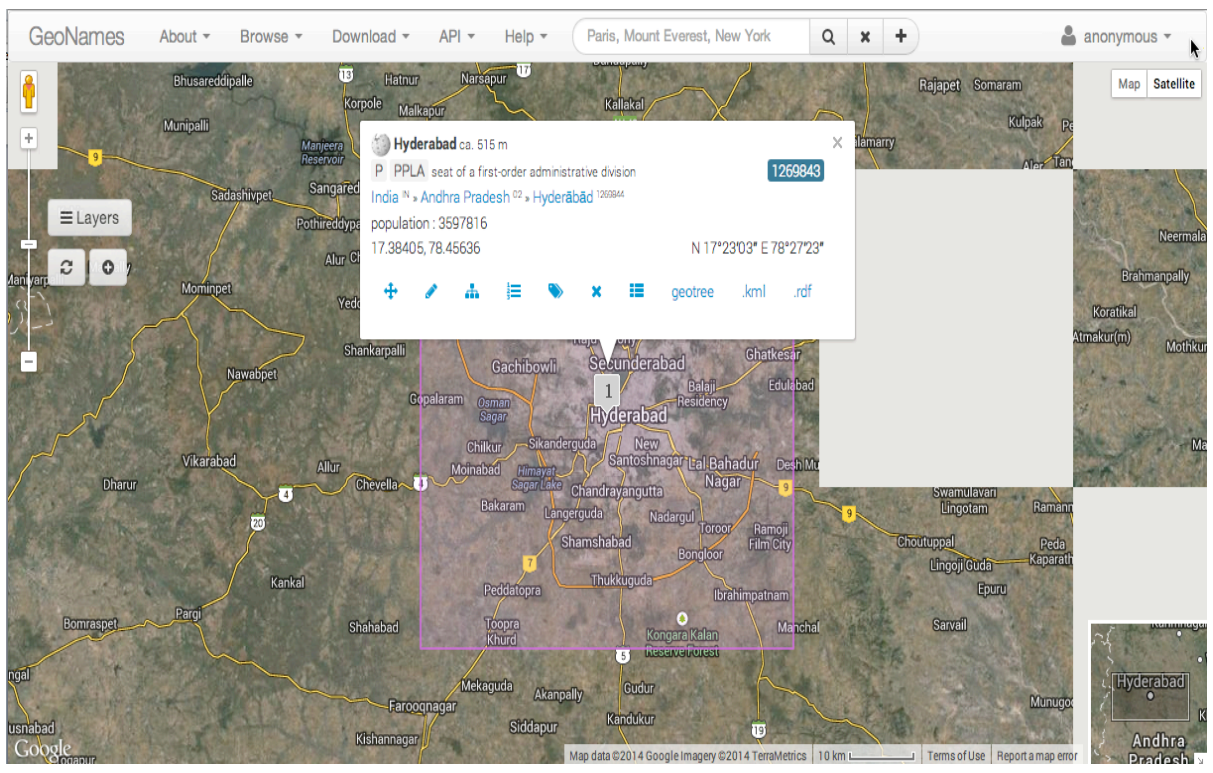

*Figure 8: GeoNames Ontology showing linked data to Hyderabad*

```xml
<?xml version="1.0" encoding="UTF-8" standalone="no"?>
<rdf:RDF xmlns:cc="http://creativecommons.org/ns#" xmlns:dcterms="http://purl.org/dc/terms/" xmlns:foaf="http://xmlns.com/foaf/0.1/"
xmlns:gn="http://www.geonames.org/ontology#" xmlns:owl="http://www.w3.org/2002/07/owl#"
xmlns:rdf="http://www.w3.org/1999/02/22-rdf-syntax-ns#" xmlns:rdfs="http://www.w3.org/2000/01/rdf-schema#"
xmlns:wgs84_pos="http://www.w3.org/2003/01/geo/wgs84_pos#">
<gn:Feature rdf:about="http://sws.geonames.org/1269843/">
<rdfs:isDefinedBy rdf:resource="http://sws.geonames.org/1269843/about.rdf"/>
<gn:name>Hyderabad</gn:name>
<gn:alternateName>Bhaganagar</gn:alternateName>
<gn:alternateName xml:lang="lt">Haidarabadas</gn:alternateName>
<gn:alternateName xml:lang="cs">Hajdarábád</gn:alternateName>
<gn:alternateName xml:lang="de">Hyderabad</gn:alternateName>
<gn:officialName xml:lang="en">Hyderabad</gn:officialName>
<gn:alternateName xml:lang="it">Hyderabad</gn:alternateName>

<gn:featureClass rdf:resource="http://www.geonames.org/ontology#P"/>
<gn:featureCode rdf:resource="http://www.geonames.org/ontology#P.PPLA"/>
<gn:countryCode>IN</gn:countryCode>
<gn:population>3597816</gn:population>
<wgs84_pos:lat>17.38405</wgs84_pos:lat>
<wgs84_pos:long>78.45636</wgs84_pos:long>
<gn:parentFeature rdf:resource="http://sws.geonames.org/1269844/"/>
<gn:parentCountry rdf:resource="http://sws.geonames.org/1269750/"/>
<gn:parentADM1 rdf:resource="http://sws.geonames.org/1278629/"/>
<gn:parentADM2 rdf:resource="http://sws.geonames.org/1269844/"/>
<gn:nearbyFeatures rdf:resource="http://sws.geonames.org/1269843/nearby.rdf"/>
<gn:locationMap rdf:resource="http://www.geonames.org/1269843/hyderabad.html"/>
<gn:wikipediaArticle rdf:resource="http://en.wikipedia.org/wiki/Hyderabad,_India"/>
<rdfs:seeAlso rdf:resource="http://dbpedia.org/resource/Hyderabad%2C_India"/>
<gn:wikipediaArticle
rdf:resource="http://ru.wikipedia.org/wiki/%D0%A5%D0%B0%D0%B9%D0%B4%D0%B0%D1%80%D0%B0%D0%B1%D0%B0%D0%B4_%28%D0%B3%D0%BE%D1%80%D0%BE%D0%
B4_%D0%B2_%D0%98%D0%BD%D0%B4%D0%B8%D0%B8%29"/>
</gn:Feature>
<foaf:Document rdf:about="http://sws.geonames.org/1269843/about.rdf">
<foaf:primaryTopic rdf:resource="http://sws.geonames.org/1269843/"/>
<cc:license rdf:resource="http://creativecommons.org/licenses/by/3.0/"/>
<cc:attributionURL rdf:resource="http://sws.geonames.org/1269843/"/>
<cc:attributionName rdf:datatype="http://www.w3.org/2001/XMLSchema#string">GeoNames</cc:attributionName>

<dcterms:created rdf:datatype="http://www.w3.org/2001/XMLSchema#date">2006-01-15</dcterms:created>
<dcterms:modified rdf:datatype="http://www.w3.org/2001/XMLSchema#date">2013-06-04</dcterms:modified>
</foaf:Document>
</rdf:RDF>
```

*Figure 9: RDF Code for the city Hyderabad in GeoNames*

## VI. CONCLUSION & FUTURE WORK

The research addresses intelligent question in Spatial Information Retrieval area. A semantic enhanced model extending the search capabilities of existing methods that is able to answer more complex queries regarding the retrieval of spatial information based on an ontology. Semantic retrieval approaches can integrate and take advantage of semantic web and information retrieval technologies and thus provide better search capabilities, achieving a qualitative improvement over keyword based retrieval by means of adopting a widely used ontology that enables information search and retrieval.

Many challenges and limitations such as the semantic referencing, heterogeneity, and the lack of an adopted ontology framework, can be pointed out as some of the main reasons for efficient spatial information retrieval. This has been addressed with Geographic ontology, which is significant in supporting the tasks of spatial information classification and organization, as clearly identified. This paper focused on application of a widely adopted GO: GeoNames, for an effective SIR, and demonstrated that the use of an ontology and inference techniques could be exploited towards an effective spatial information retrieval. Moreover, future SIR System will be more intelligent and be able to answer questions in natural language description, which will address the quality of spatial information retrieval across various digital libraries.